\documentclass[reprint,amsmath,amssymb,nofootinbib,aps,prd,superscriptaddress,twocolumn,RNAAS,numberedappendix]{aastex631}
\usepackage{aas_macros}
\usepackage{graphicx}
\usepackage{dcolumn}
\usepackage{bm}
\usepackage{comment}
\usepackage{amsmath}
\usepackage{float}
\usepackage{lipsum}
\usepackage{color}
\usepackage{times}
\usepackage{textcomp}
\usepackage{latexsym}
\usepackage{acro}
\usepackage{mathtools}
\usepackage{url}
\usepackage[utf8]{inputenc}
\usepackage{physics}
\usepackage{natbib}
\usepackage{soul}
\setcitestyle{authoryear,open={(},close={)}}

\begin{document}

\title{Strong lensing: A magnifying glass to detect gravitational-wave microlensing}
\date{\today} 

\author{Eungwang Seo}
\thanks{E-mail: ewseo@phy.cuhk.edu.hk}
\affiliation{Department of Physics, The Chinese University of Hong Kong, Shatin, N.T., Hong Kong}
\author{Otto A. Hannuksela}
\affiliation{Department of Physics, The Chinese University of Hong Kong, Shatin, N.T., Hong Kong}
\author{Tjonnie G. F. Li}
\affiliation{Department of Physics, The Chinese University of Hong Kong, Shatin, N.T., Hong Kong}
\affiliation{Department of Physics and Astronomy, Katholieke Universiteit Leuven, Celestijnenlaan 200D box 2415, 3001 Leuven, Belgium}
\begin{abstract}
    Microlensing imprints by typical stellar-mass lenses on gravitational waves are challenging to identify in the LIGO--Virgo frequency band because such effects are weak. However, stellar-mass lenses are generally embedded in lens galaxies such that strong lensing accompanies microlensing. Therefore, events that are strongly lensed in addition to being microlensed may significantly improve the inference of the latter. We show a proof-of-principle demonstration of how one can use parameter estimation results from one strongly lensed signal to enhance the inference of the microlensing effects of the other signal with the Bayesian inference method currently used in gravitational-wave astronomy.
    We expect this to significantly enhance our future ability to detect the weak imprints from stellar-mass objects on gravitational-wave signals from colliding compact objects.
    
\end{abstract}
\section{Introduction}
\begin{figure*}
 {\includegraphics[width=0.49\textwidth]
  {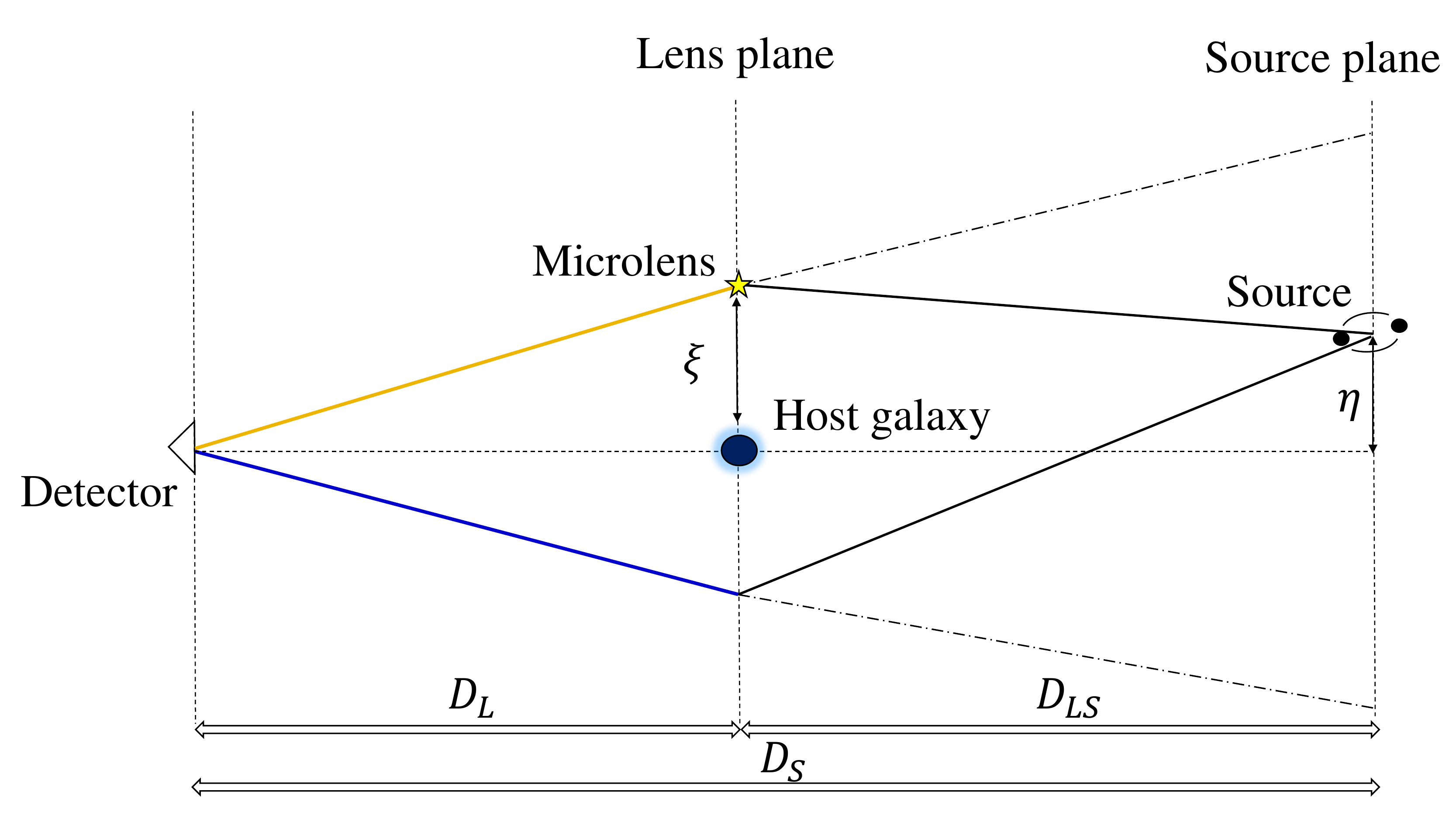}}
 {\includegraphics[width=0.49\textwidth]
  {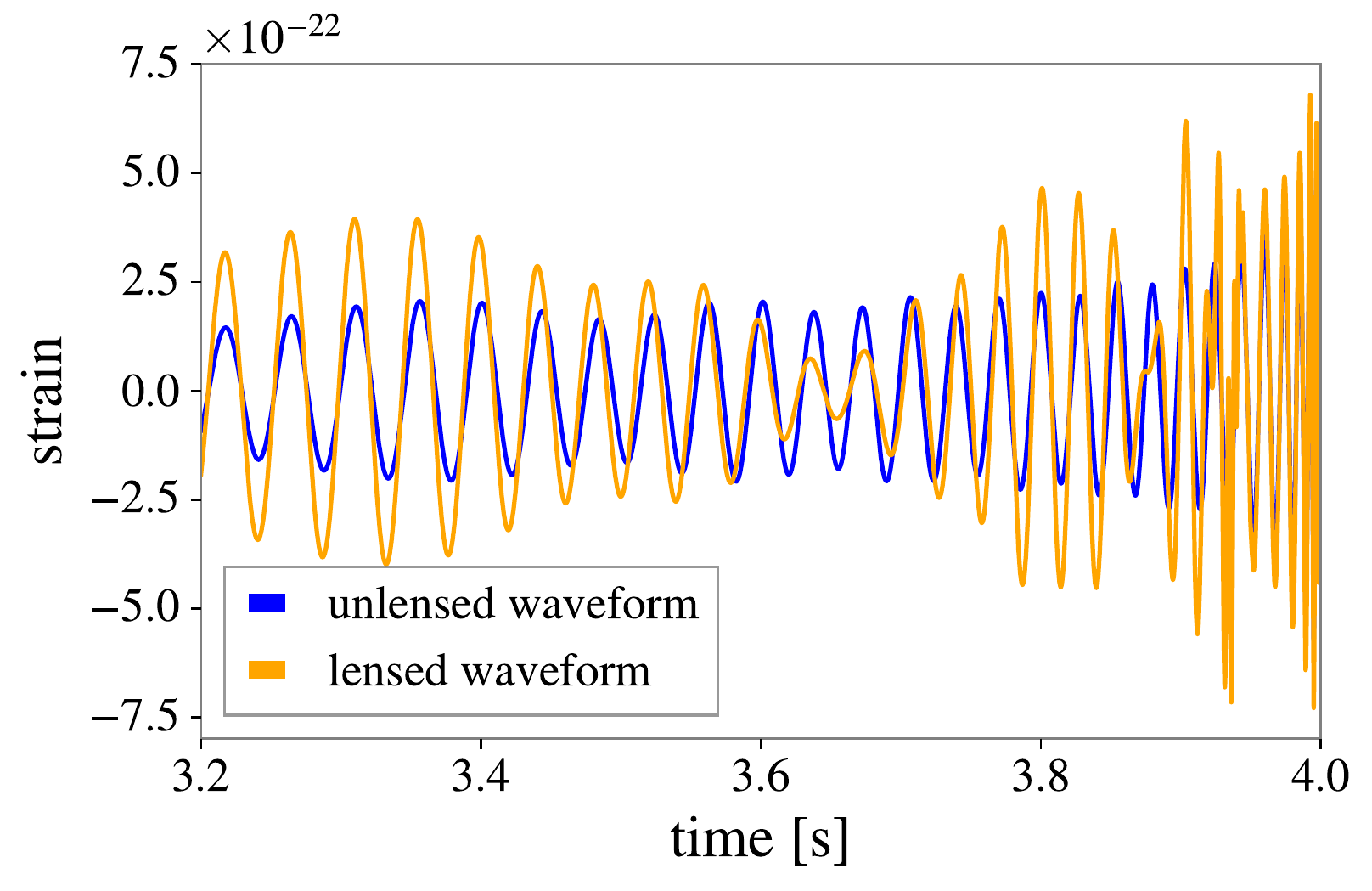}}
 \caption{\textit{Left panel}: GWs from a source propagates near a lens galaxy of microlens (yellow star). 
One of the signals undergoes microlensing and strong lensing (orange-solid line), and the other one undergoes only strong lensing by the lens galaxy (blue-solid line).
$D_L$, $D_S$ and $D_{LS}$ is the angular diameter distance between the detector and the lens galaxy, the detector and the source, and the lens galaxy and the source, respectively. 
\textit{Right panel}: A mock microlensed (orange) and unlensed (blue) GW signal from a binary black hole in the time domain segment.
Since the GW propagates near a microlens, the signal is distorted.
For this figure, the microlens parameters are $M^{z}_{\textrm{ML}}=3000\, \rm M_{\odot}$ and $y=0.5$ to show an example of appreciably visible beating patterns. 
Note that the unlensed GW signal appears to have small beating patterns because the waveform approximant is \textsc{IMRPhenomPv2} which involves spin precessions.}
\label{fig:figure-one}
\end{figure*}
A gravitational lens is a massive object that distorts spacetime, which deflects light rays from a distant source toward an observer~\citep{meylan2006gravitational}.
In a similar manner to light, a gravitational wave (GW) is deflected when it passes by a massive object~\citep{wang1996gravitational, nakamura1998gravitational, takahashi2003wave}.
If the lens is at a galaxy scale, one can observe multiple signals from one GW source, so-called strong lensing~\citep{10.1093/mnras/sty2145,10.1093/mnras/sty411,PhysRevD.97.023012}
On the other hand, microlensing is caused by a lens whose mass is between $[1,10^{5}]M_{\odot}$ that generates a shorter time delay than a chirp time of a GW signal.
When a GW travels near a microlens, the microlens can imprint "beating patterns" on the GW, which could potentially lead to microlensing observations by the current gravitational-wave detectors~\citep{PhysRevD.98.103022,PhysRevLett.122.041103,PhysRevD.90.062003,refId0,PhysRevD.101.123512}.
Unfortunately, the microlensing signatures, particularly due to stellar-mass objects, are weak~\citep{10.1093/mnras/stab579,mishra2021gravitational}. 
Thus, it is often challenging to obtain compelling evidence of such signatures.

Typical stellar-mass objects are predominantly part of larger-scale structures, which may induce strong lensing effects on GWs. 
Therefore, when strong lensing occurs, it is realistic that the effect could be observed in conjunction with microlensing signatures.
In particular, suppose that a GW from a distant black hole binary undergoes strong lensing by a galaxy hosting microlenses. 
The GW is then split into two or more signals arriving minutes to months apart~\citep{10.1093/mnras/sty2145,10.1093/mnras/sty411,PhysRevD.97.023012,Liu_2021,lo2021bayesian,janquart2021fast,theligoscientificcollaboration2021search}, where each signal may include different beating patterns caused by the field of stellar-mass microlenses along their respective paths (Fig.~\ref{fig:figure-one}; see also Refs.~\citep{10.1093/mnras/stab579,mishra2021gravitational}.
In addition to the usual beating patterns, the strong lensing galaxy could even further amplify the microlensing effect~\citep{Dai:2016igl,refId0,meena2020gravitational,pagano2020lensinggw,10.1093/mnras/stab579,mishra2021gravitational}.

In this letter, we focus on the question: "How  does incorporating  strong lensing signals improve our ability to detect even very faint microlensing effects in parameter estimation?"
We will obtain information from the first strongly lensed signal by conducting parameter estimation (PE) to analyze the second signal, which is both microlensed and strongly lensed. 
Finally, we will show that we can detect the microlensing effects on the second signal at an improved accuracy.
\section{Methodology}
In principle, one or both of the strongly lensed gravitational waves in Fig.~\ref{fig:figure-one} can undergo non-negligible microlensing.
We consider two cases where either only one signal or both signals are detected as being microlensed with current detectors sensitivities~\citep{schechter2002quasar, dobler2006microlensing, vernardos2019quasar, list2020sensitivity}.
The reason is that the probability that only one of the signals is considerably microlensed to a detectable degree is higher than the probability that both of them are, except in some select situations in which the stellar density is high enough to compensate for generally uncommon microlensing occurrence or extreme magnifications are involved~\citep{refId0,meena2020gravitational}.

Therefore, firstly, suppose that the first signal (signal 1) undergoes microlensing by a microlens embedded in the lens galaxy, and the other one (signal 2) does negligible microlensing (see Fig.~\ref{fig:figure-one}, for an illustration).
In that case, we can carry out PE for signal 2 and, since the two strongly lensed signals are related~\citep{Liu_2021,lo2021bayesian,janquart2021fast,theligoscientificcollaboration2021search}, use the results to reduce degeneracies in the parameter estimation of signal 1.
In other words, the waveform of signal 1 can be retrieved by conducting PE on signal 2.
Note, however, that the apparent luminosity distances ($d_L$), coalescence times ($t_c$), and coalescence phases ($\phi_c$) can differ between the two signals~\citep{haris2018identifying,dai2020search,wierda2021detector,janquart2021fast}, so we take them as free parameters.
Furthermore, we neglect the effect of higher-order modes on the waveform, which could induce minor additional signatures~\citep{janquart2021identification}.

Secondly, suppose that both signal 1 and signal 2 are microlensed.
In this scenario, one may retrieve a wrong template from the PE for signal 2 if the previous scenario is assumed.
Thus, microlensed templates should be applied to the PE for signal 2 to get the more accurate retrieved waveform.
However, in the same condition, the retrieved waveform of signal 1 from the PE results of signal 2 inevitably has lower Bayes factors and broader posteriors than ones in the previous scenario because signal 2 has more parameters (microlens parameters are added) and degeneracies.

The microlensed waveform $h_{\rm ML}$ in the frequency domain is
\begin{equation}
\label{EQ1}
    h_{\rm ML}(f,\theta_{\rm s},\theta_{\rm ML})=F(f,M^{z}_{\rm ML},y)\times h_{\rm UML}(f,\theta_{\rm s}),
\end{equation}
where $\theta_{\rm s}$ is source parameters, $\theta_{\rm ML}$ is microlens parameters $\{M^{z}_{\rm ML},y\}$, and $F(f)$ is the amplification factor in wave optics~\citep{takahashi2003wave}.
Source parameters ($\theta_{\rm s}$) are classified into intrinsic parameters ($\theta_{\rm int}$): Component masses and 3-dimensional component spins, and extrinsic parameters ($\theta_{\rm ext}$) including sky position, inclination and polarization.
We use the \textsc{IMRPhenomPv2} approximant, which includes inspiral, merger, and ringdown phase of precessing binary black holes~\citep{hannam2014simple} for unlensed GW signal $h_{\rm UML}$.
Similar to previous microlensing analyses~\citep{hannuksela2019search,theligoscientificcollaboration2021search}, we choose the isolated point mass lens model for our microlens.
The amplification factor ($F(f)$) of this model is
\begin{align}
\label{EQ2}
    F(w)&=\exp\bigg{[}\dfrac{\pi w}{4}+i \dfrac{w}{2} \bigg{\{}\text{ln} \bigg{(}\dfrac{w}{2}\bigg{)}-2\phi_m(y)\bigg{\}}\bigg{]} \nonumber \\
&\times \Gamma \bigg{(}1- \dfrac{i}{2}w\bigg{)} {_1}F_1\bigg{(}\dfrac{i}{2}w, 1; \dfrac{i}{2}w y^2\bigg{)}
\end{align}

To conduct PE, We consider three hypotheses. 
Firstly, the hypothesis $\mathcal{H}_{\rm SL+ML}$ means that a GW signal (primary signal) is the one part of strongly lensed signals split by a lens galaxy and is also microlensed by a point mass.
The other parts of the strongly lensed signals (auxiliary signals) are counterparts of the primary signal, and thus all signals have the same intrinsic parameters ($\theta_{\rm int}$), but each signal has its own unique $d_L$, $t_c$ and $\phi_c$.
The hypothesis $\mathcal{H}_{\rm SL+ML}$ needs an assumption that not less than one of the auxiliary GW signals are detected so that one can obtain parameter information from the auxiliary signals to fix the source parameters of the primary signal.

To be specific, we choose the maximum-likelihood waveform to obtain a reference waveform from the PE results of the auxiliary signals (See e.g. \cite{dai2020search})\footnote{This paper is the first paper to perform a joint likelihood computation using the maximum-likelihood waveform. We adopted a similar approach to fix the source parameters of signal 1. (See Eq. A5 in~\citep{dai2020search})}.
The reference waveform only includes a single value of source parameters ($\theta_{\rm s}$) from the maximum likelihood estimated by the PE except for $d_L$, $t_c$, and $\phi_c$.
As was mentioned before, strong lensing endows lensed signals with respective magnifications, time delays and phase shifts, and thus $d_L$, $t_c$, and $\phi_c$ are unavailable for the reference waveform.

Secondly, the hypothesis $\mathcal{H}_{\rm ML}$ means that a GW signal is microlensed by an isolated point mass lens, which is also assumed in~\citep{hannuksela2019search,theligoscientificcollaboration2021search}.
Microlensed waveforms under the hypothesis $\mathcal{H}_{\rm ML}$ are fully determined by Eq.\ref{EQ1} and Eq.\ref{EQ2}.
Conversely, the unmicrolensing hypothesis $\mathcal{H}_{\rm UML}$ means that a GW signal is unmicrolensed, which is opposite of the $\mathcal{H}_{\rm ML}$.
We compare the PE results assuming the $\mathcal{H}_{\rm ML}$ and the $\mathcal{H}_{\rm SL+ML}$ in the microlensing inference and show how incorporating strong lensing allows us to constrain the microlens parameters better.

In practice, the strong lensing would amplify the microlensing effect (particularly for Type-I images)~\citep{refId0,meena2020gravitational,pagano2020lensinggw,10.1093/mnras/stab579,mishra2021gravitational}, further boosting our ability to detect the microlensing signatures, namely, beating patterns.
Also, shear in a strong lens would likely enhance the magnification of a microlensed signal~\citep{huterer2005effects}, which makes the beating patterns more prominent.
However, modeling these more realistic microlensing scenarios would also increase the parameter space required in the inference.
In turn, this may hinder our ability to discern the precise parameters of the individual microlens.

To investigate the interaction between strong lensing and microlensing by PE, one needs a more complicated lens model using full-wave optics.
However, the current PE studies are still limited to the isolated point mass lens~\citep{PhysRevD.98.083005,hannuksela2019search,basak2021constraints,chung2021lensing,theligoscientificcollaboration2021search}, due to computational limitations.
Also, efforts to develop accurate microlensed waveforms capturing such "macromodel" (galaxy lens being the macromodel in our example) effects are currently underway~\citep{Dai:2018enj,pagano2020lensinggw,10.1093/mnras/stab579}.
We leave investigating the interplay between the increased microlensing parameter space and the boost from the strong lensing enhancement for future work.

We adopt \textsc{bilby} \citep{Ashton_2019} to carry out PE and use \textsc{dynesty} \citep{10.1093/mnras/staa278} for the nested sampler.
For each injection, we select the default prior setting for a precessing binary black hole in \textsc{bilby} and assume design sensitivity  \citep{PhysRevX.11.021053,abbott2020prospects,PhysRevLett.123.231108}.
In addition, we choose a uniform distribution for redshifted lens mass prior and set the upper limit as $150\, \rm M_{\odot}$. 
The source position prior is proportional to normalized $y$ and has range between [0, 3] (as in~\citep{PhysRevD.98.083005,theligoscientificcollaboration2021search}) for injected, redshifted lens masses $M^{z}_{\rm ML} \geq 20\, \rm M_{\odot}$. 
For lighter lenses, the microlensing effect is too weak to be detected beyond the Einstein radius of the microlens, so we set the normalized $y$ range as [0, 1].
We focus on the lighter lenses because beating patterns caused by them are usually hidden below the current detector noise level.
It is hard to discern their presence with strong evidence if one only considers the microlensing effects from a point mass lens~\citep{hannuksela2019search,theligoscientificcollaboration2021search}. 
Also, lighter lenses are more common to occur microlensing than heavier lenses~\citep{refId0,mishra2021gravitational}.
\section{Results}
\begin{figure}[t]
\includegraphics
  [width=1.0\hsize]
  {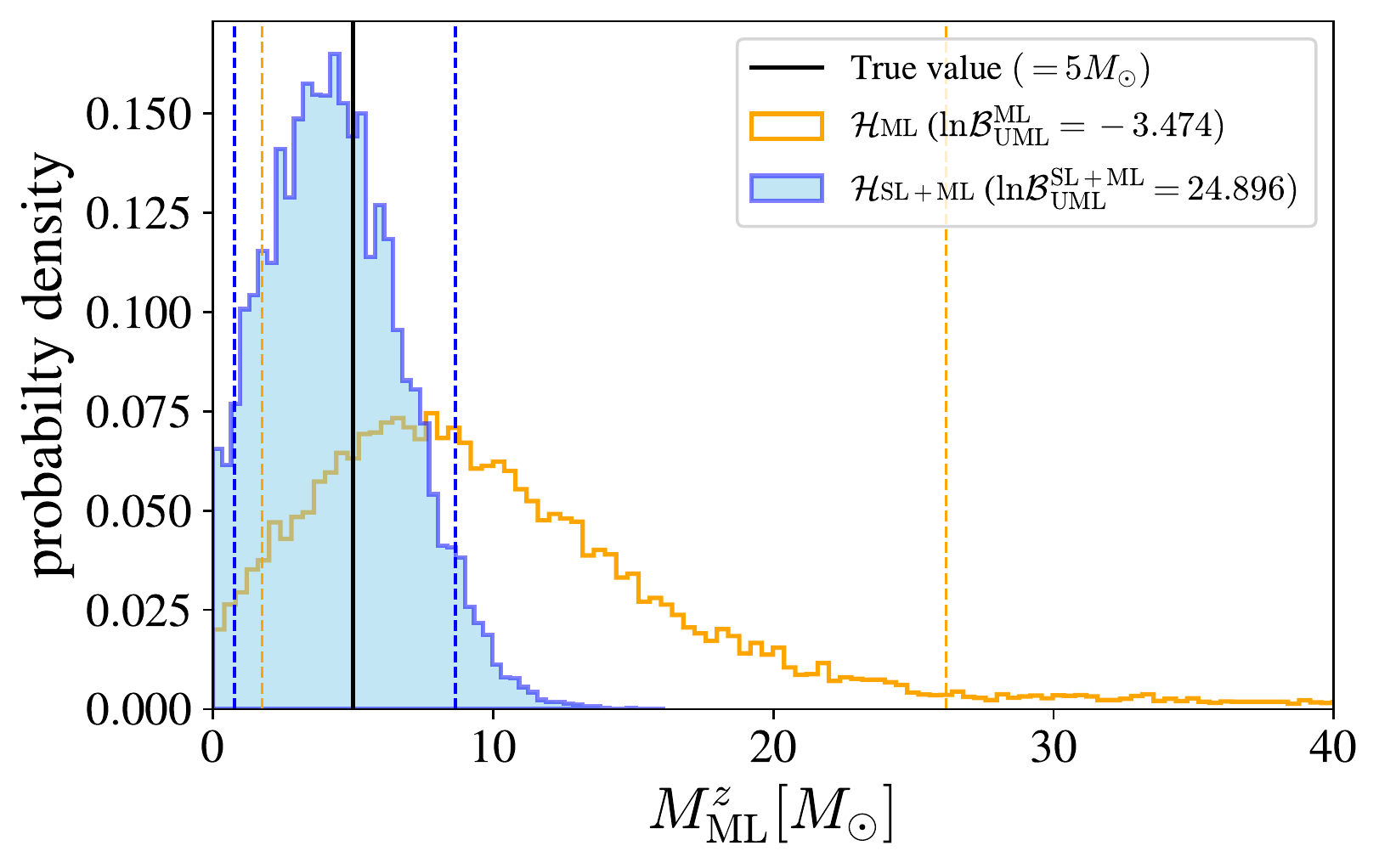}
\caption{1D posterior probability distributions of redshifted lens mass marginalized over source position. 
The densities for the $5\, \rm M_{\odot}$ microlens injections are estimated depending on whether the GW undergoes strong lensing or not.
The blue shaded and the orange lined histogram show the posterior distribution with the $\mathcal{H}_{\rm SL+ML}$ and the $\mathcal{H}_{\rm ML}$, respectively. 
The vertical black solid line marks the true value, and each colored dashed lines mark its 90\% credible intervals.
The posterior of the redshifted lens mass of the microlens (blue) converges well to the true value for the signal undergoing strong lensing, and the Bayes factor favors microlensing.
On the other hand, the posterior (orange) for the signal not undergoing strong lensing is more spread in the prior range, and the Bayes factor is negative.}
\label{fig:two-column-figure}
\end{figure}
\begin{figure*}
 {\includegraphics[width=1.0\textwidth]
  {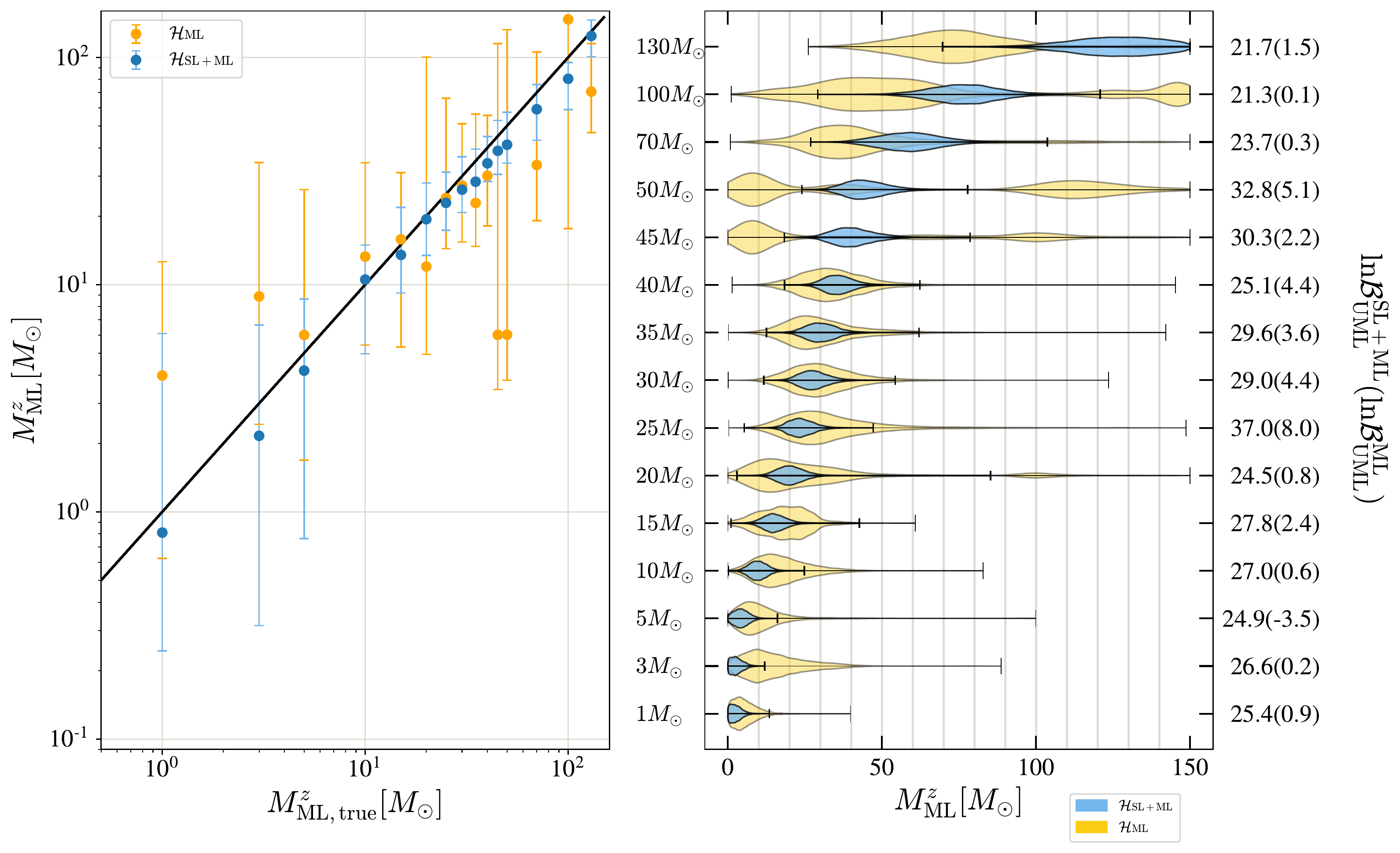}}
	\caption{\textit{Left panel}: The maximum posterior probability points (dots) and 90$\%$ credible intervals (whiskers) of the inferred redshifted lens mass as a function of the true redshifted lens mass (shown on a logarithmic scale for clarity).
	The x-axis is the true value of redshifted lens mass, and the y-axis is the estimated value of injections.
    Whether incorporating the strong lensing hypothesis or not is expressed by blue and orange color, respectively.
    The black solid line shows the true value.
    The results incorporating strong lensing (blue dots) recover the true value with greater accuracy.
	\textit{Right panel}: 
	The violin plots show the 1D marginalized posteriors of the 15 injected redshifted lens mass over normalized source position for both cases with (blue) and without (orange) strong lensing.
	The true value of each injection is shown on the left-hand side of the y-axis. 
	The Bayes factor with and without incorporating strong lensing is shown on the right axis. 
	The microlens mass is better recovered when incorporating strong lensing.}
\label{fig:third-one}
\end{figure*}
\subsection{Scenario A: One of two signal is microlensed.}
\label{scenarioA}
As an illustrative example to inspect the role of strong lensing in the microlensing analysis, we simulate two binary black holes (BBH) with masses ($30 \rm M_{\odot}, 30 \rm M_{\odot}$) lensed by a stellar-mass lens $M^{z}_{\rm ML}=5\, \rm M_\odot$. 
The first BBH event, which is identified as a single lensed signal, is at redshift $z_s=0.2$, and the redshift of the microlens is $z_l=0.1$.
In contrast, the second BBH event consists of two signals (signal 1 and signal 2) due to strong lensing effects by the lens galaxy of the stellar-mass microlens.
Signal 1 is microlensed by the stellar-mass lens, but signal 2 is not.
For a fair comparison, we tune $z_s$ and $z_l$ of the second event so that the signal-to-noise ratios (SNR) of the two BBH events are the same (The strong lensing can magnify the second event).

We find that the second event, which is both  microlensed and strongly lensed, is detected at a Bayes factor of $\ln{{\mathcal{B}^{\rm SL+ML}_{\rm UML}}}\sim 24.9$, 
while the first event that did not undergo strong lensing has weak evidence in favor of the $\mathcal{H}_{\rm ML}$ ($\ln{{\mathcal{B}^{\rm ML}_{\rm UML}}}\sim -3.5$).
$\mathcal{B}^{\rm SL+ML}_{\rm UML}$ is the Bayes factor between the $\mathcal{H}_{\rm SL+ML}$ and the $\mathcal{H}_{\rm UML}$, while $\mathcal{B}^{\rm ML}_{\rm UML}$ is the Bayes factor between the $\mathcal{H}_{\rm ML}$ and the $\mathcal{H}_{\rm UML}$, which is the same as the Bayes factor defined in ~\citep{theligoscientificcollaboration2021search}.
In Fig.~\ref{fig:two-column-figure}, we display the 1D marginalized posteriors of the redshifted lens mass of two events lensed by the same microlens but assuming a different hypothesis ($\mathcal{H}_{\rm SL+ML}$ or $\mathcal{H}_{\rm ML}$) with the corresponding Bayes factor.
In the case that the event undergoes strong lensing, the redshifted lens mass is well-recovered.
In contrast, the posterior for the microlensed event by an isolated point mass is recovered less accurately. 
This example shows that detecting multiple signals from a GW source can significantly improve the microlensing search.

We also simulate more mock signals which have SNR $\sim$ 20 lensed by microlens with various masses from $1\, \rm M_{\odot} \sim 150\, \rm M_{\odot}$ and fixed source position ($y=0.5$).
The left panel of Fig.~\ref{fig:third-one} shows the estimated posterior ranges (90$\%$ credible intervals) with dots indicating the maximum posterior probability of redshifted lens mass under the $\mathcal{H}_{\rm ML}$ and $\mathcal{H}_{\rm SL+ML}$. 
Meanwhile, violin plots in the right panel show the posteriors of the redshifted lens mass of two cases. 
The posteriors of the redshifted lens mass are better constrained to the true values when the strong lensing is applied to microlensing analysis. 
In addition, high Bayes factors show strong evidence that the events are microlensed.
Note that the redshifted lens mass recovery becomes weaker for higher injected lens masses due to degeneracies between the normalized source position $y$ and the redshifted lens mass $M^z_{\rm ML}$.

Conversely, the PE results under the $\mathcal{H}_{\rm ML}$ show biases towards higher or lower masses for each injection, and the posteriors for some high lens mass injections have multiple peaks (orange plots).
The posteriors are broader than ones under $\mathcal{H}_{\rm SL+ML}$ for the lower mass injections because degeneracies between the source parameters can imitate the microlensing effects. 
Furthermore, more than half of the events have low Bayes factor ($\ln{{\mathcal{B}^{\rm ML}_{\rm UML}}} < 1.7$) which is in statistical fluctuations expected for unlensed events~\citep{theligoscientificcollaboration2021search}. 
Therefore, the estimated Bayes factors included in the fluctuation range indicate no microlensing effects in our simulations.
\subsection{Scenario B: Both signals are microlensed}
\label{scenarioB}
Since microlens candidates are pervasive in strong lensing regions of a lens galaxy, both two strongly lensed signals (signal 1 and signal 2) could also be microlensed (by $\rm ML_{1}$ and $\rm ML_{2}$). 
Therefore, we simulate two binary black holes with the same parameters as the above example in Fig.~\ref{fig:two-column-figure}, but $M^{z}_{\rm ML_{1}}=30\, \rm M_\odot$, and put another microlens ($\rm ML_{2}$) on the path of signal 2.
For the microlens masses, we adopt the values $M^{z}_{\rm ML_{2}}=5,\, 30,\, 50\, \rm M_\odot$.
According to the normal PE results under the $\mathcal{H}_{\rm ML}$ (Fig.~\ref{fig:third-one}, orange color), the recovered posteriors have not converged well; the retrieved maximum likelihood waveform can be highly biased if the newly introduced microlens is above $40M_{\odot}$.
We find that if the microlens on the path of signal 2 is not massive $(M^{z}_{\rm ML_{2}} = 5 M_{\odot} \,  (30M_{\odot}))$, signal 1 is detected at a Bayes factor of $\ln{{\mathcal{B}^{\rm SL+ML}_{\rm UML}}}\sim 20 \, (15)$ with well-recovered posteriors.
\begin{figure}[hbt!]
\includegraphics
[width=1.0\hsize]
{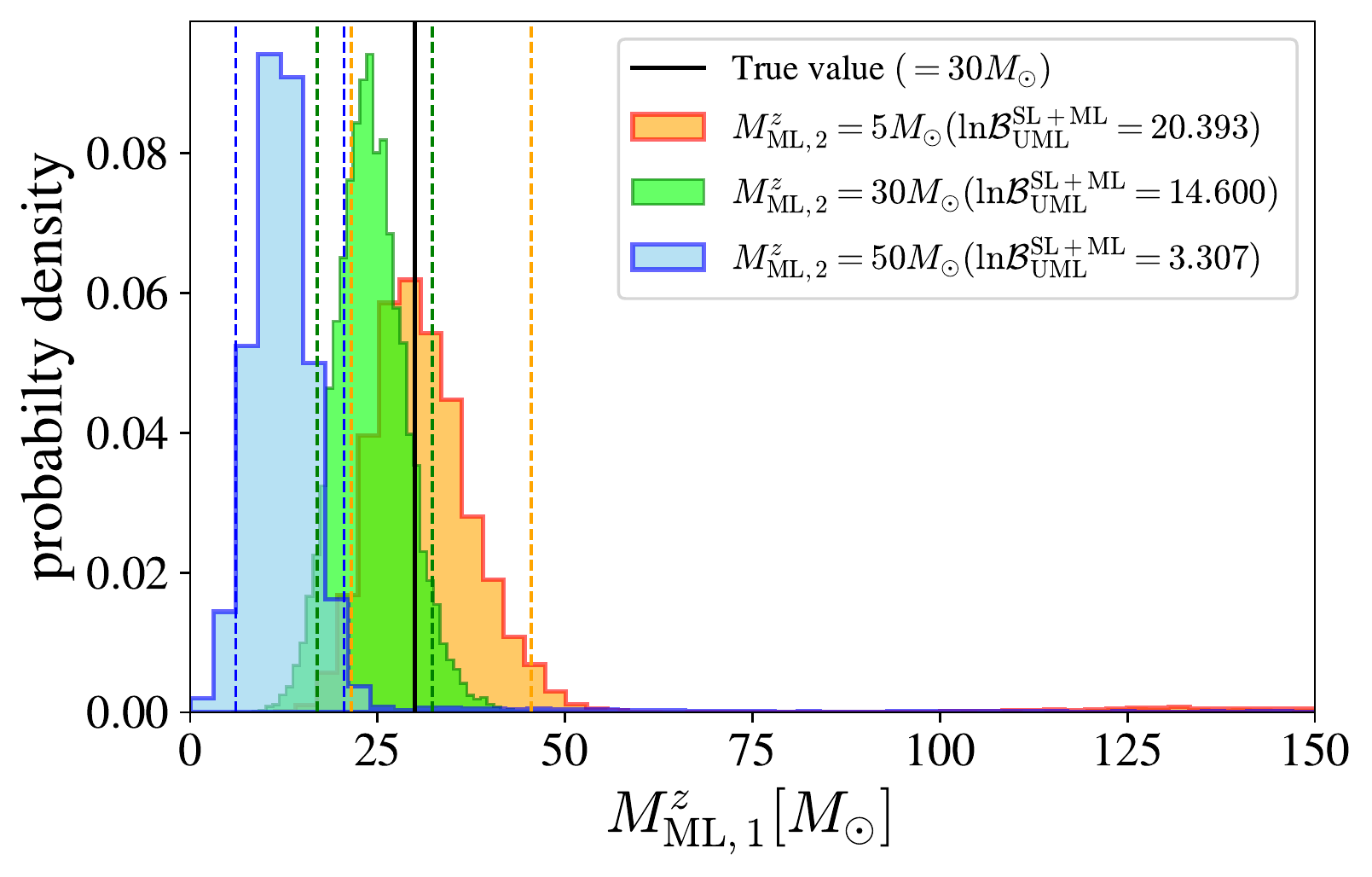}
\caption{1D marginalized posterior probability distributions of $M^{z}_{\rm ML_{1}}$ ($30\, \rm M_{\odot}$).
Here both images undergo strong lensing, but signal 2 is also microlensed.
Orange, green and blue colored histograms indicate the case that $M^{z}_{\rm ML_{2}} = 5, \, 30$ and $50 \, \rm M_{\odot}$, respectively.
The vertical black solid line marks the true value, and each colored dashed lines mark the 90\% credible intervals.
The inference of the heavier $M^{z}_{\rm ML_{2}}$ are biased towards lower value and have lower Bayes factors.
}
\label{fig:fourth-column-figure}
\end{figure}
\begin{figure}[!t]
\includegraphics
[width=1.0\hsize]
{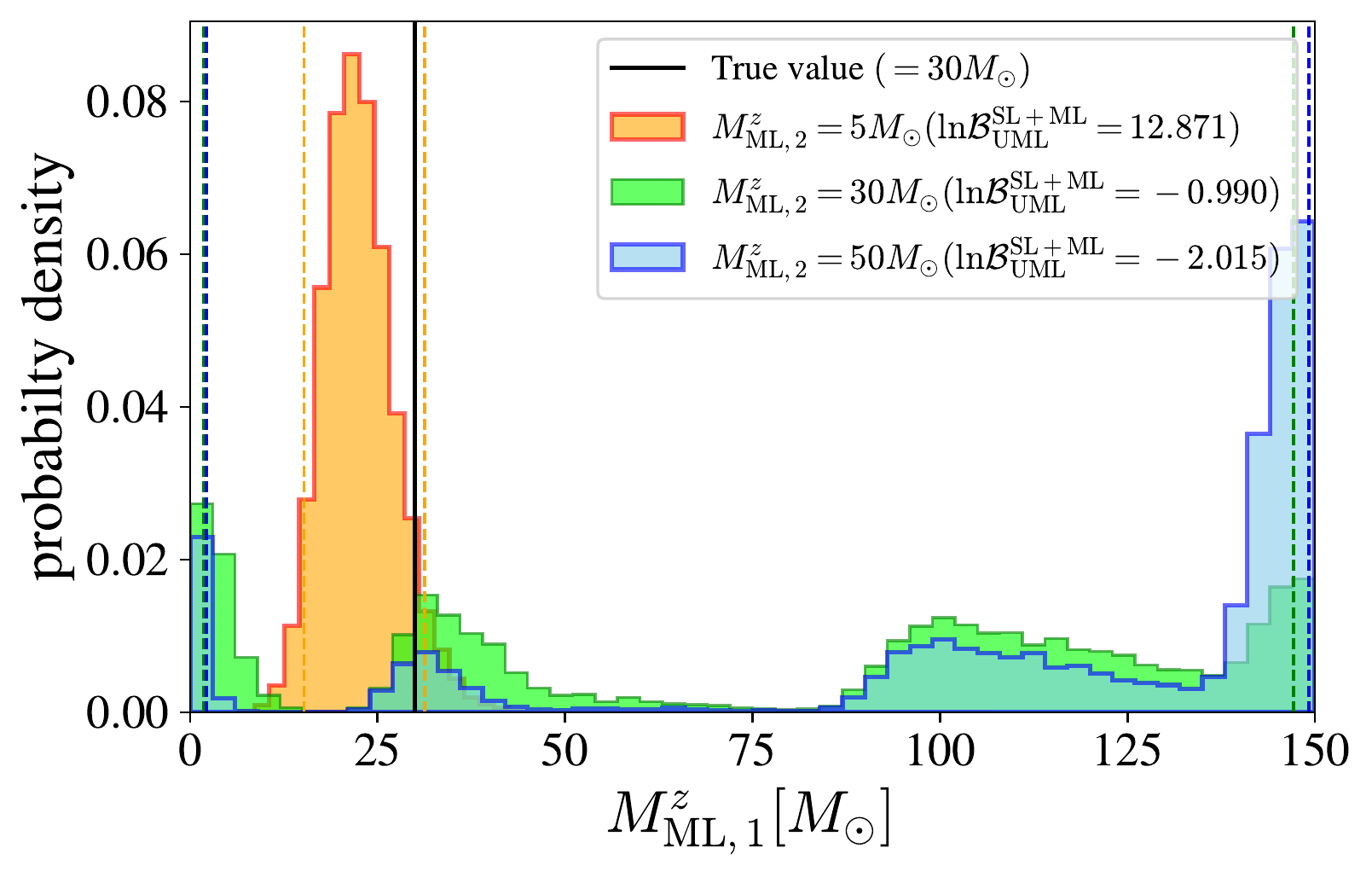}
\caption{Same configurations as in Fig.~\ref{fig:fourth-column-figure}, but with the hypothesis that GWs undergo strong lensing and signal 2 is not microlensed.
For the $30M_{\odot}$ and $50M_{\odot}$ cases, the posteriors have not converged with negative Bayes factors. 
Only the $5M_{\odot}$ case has a peak due to a relatively weak lensing effect, but it is at the lower value compared with the $5M_{\odot}$ case in Fig.~\ref{fig:fourth-column-figure}.}
\label{fig:fifth-column-figure_UT}
\end{figure}

In Fig.~\ref{fig:fourth-column-figure}, we show the 1D marginalized posteriors of the redshifted lens mass of three events lensed by same microlens $M^{z}_{\rm ML_{1}}=30\, \rm M_\odot$ but the masses of $\rm ML_{2}$ are different.
Note that $\mathcal{H}_{\rm ML}$ is assumed in the inference of signal 2.
For all three cases, each posterior peaks below the true value.
In addition, the accuracy decreases further when the lensing effect is more substantial.

On the other hand, in Fig.~\ref{fig:fifth-column-figure_UT}, we show the same posteriors, but $\mathcal{H}_{\rm UML}$ is used on signal 2. 
As mentioned above, using the wrong hypothesis retrieves the wrong waveform templates from signal 2, which biases the PE results for signal 1.
The estimated posteriors are far from the true value for heavy mass cases.
Only $M^{z}_{\rm ML_{2}}=5M_{\odot}$ case has a peak because the lensing effect on signal 2 is not that significant, but it is converged to lower $M^{z}_{\rm ML_{1}}$ compared to the posterior of the $5M_{\odot}$ case in Fig.~\ref{fig:fourth-column-figure}.
Comparing the posteriors estimated under different hypotheses enables us to identify microlensing effects on signal 2, especially when the second microlens is heavier than a few tens of solar masses.
\section{Conclusion}
GWs lensed by typical stellar-mass objects are difficult to detect at the current LIGO-Virgo sensitivities.
Despite this, detecting and confirming microlensed GWs is valuable to study the substructures of the lens galaxy, including the population of stars and compact objects.

Stellar-mass microlenses are principally embedded in their lens galaxy, and it is plausible for GWs to undergo both strong lensing and microlensing effects.
In this context, we have shown that microlensed GWs could be detected with more considerable statistical significance by utilizing auxiliary signals from strong lensing compared to solely microlensed GWs, and the mass of the lens is well-recovered.
Indeed, the degeneracies in the microlensed GWs can be reduced by fixing their source parameters.
Since source parameters are common for all strongly lensed signals (except for luminosity distance, coalescence time, and coalescence phase), one can use the maximum likelihood waveform retrieved from one signal to infer parameters of the others (Fig.~\ref{fig:third-one}).
By doing so, one can constrain the lens parameters of the microlens with improved accuracy.

Consequently, we could detect even low-mass microlenses of a few solar masses with great accuracy, although this is not possible in the absence of strong lensing.
Also, sub-solar mass microlenses could perhaps be detectable with next-generation detectors. 
Such findings are particularly essential for the study of lensing statistics and the detection of primordial black holes.

In an actual search, we would not know which (if any) of the events are microlensed. 
Thus, we would perform the analysis assuming that the first, second, or both of the images are lensed separately, choosing the one with the best supporting evidence.
Therefore, we have considered the scenarios where one or both of the signals are microlensed and shown that the improvement is apparent in both cases.
As discussed previously, the probability that both strongly lensed signals are considerably microlensed is lower than the probability that one is, except when dealing with very high stellar densities or extreme magnifications~\citep{refId0}.

In this work, we have neglected the impact of the lens galaxy on the microlens.
The lens galaxy amplifies the microlensing effect, and the morphology of the microlensed waveform will be more complicated.
While the effect of the lens galaxy on the microlens has been investigated in the context of wave optics approximations in the past~\citep{refId0,meena2020gravitational,10.1093/mnras/stab579,mishra2021gravitational}, we can not utilize it in parameter estimation due to computational reasons.
However, if a fast algorithm to conduct such a complex PE using full-wave optics is developed in the future, one can apply more complex waveform models to a PE on microlensed signals.
Of course, resolving the individual microlensed parameters could become more difficult due to the increased parameter space of the more complex waveform, for example, due to degeneracies between waveform parameters. 
Nevertheless, we expect assuming the hypothesis $\mathcal{H}_{\rm SL+ML}$ enhances the PE for microlensed signals, which allows us to find more substantial evidence for microlensing than assuming the hypothesis $\mathcal{H}_{\rm ML}$ if the waveform used in the PE is similar to the true waveform.
In fact, the lens galaxy could greatly magnify a GW signal and increase the SNR.
Typically, PE shows smaller error ranges on posteriors of source parameters for high SNR events than low SNR events, and thus we can get a more accurate reference waveform to fix source parameters of a microlensed signal.
\section{acknowledgement}
The work described in this paper was partially supported by grants from the Research Grants Council of the Hong Kong (Project  No. CUHK  24304317),  The  Croucher Foundation of Hong Kong, and the Research Committee of the Chinese University of Hong Kong, and the research program of the Netherlands Organisation for Scientific Research (NWO). 
We would like to thank all participants of the LVC lensing group for helpful discussions; A. Ganguly for a sincere review.
We are grateful for computational resources provided by the LIGO Laboratory and supported by the National Science Foundation Grants PHY-0757058 and PHY-0823459.
\newpage
\bibliographystyle{aasjournal}
\bibliography{main}
\end{document}